\documentclass[10pt, conference]{IEEEtran}
\IEEEoverridecommandlockouts
\usepackage{cite}
\usepackage{amsmath,amssymb,amsfonts}
\usepackage{algorithmic}
\usepackage{graphicx}
\usepackage{textcomp}
\usepackage{xcolor}
\usepackage{listings}
\usepackage{gensymb}
\usepackage{array}
\def\BibTeX{{\rm B\kern-.05em{\sc i\kern-.025em b}\kern-.08em
    T\kern-.1667em\lower.7ex\hbox{E}\kern-.125emX}}

\renewcommand{\baselinestretch}{0.940}

\begin{document}

%\title{DORY MIXED - DATE 2023}
\title{A 3 TOPS/W RISC-V Parallel Cluster for Inference of Fine-Grain Mixed-Precision Quantized Neural Networks}

%\author{Authors omitted for double-blind review}
\author{\IEEEauthorblockN{
        Alessandro Nadalini\IEEEauthorrefmark{1},
        Georg Rutishauser\IEEEauthorrefmark{2},
        Alessio Burrello\IEEEauthorrefmark{1},
        Nazareno Bruschi\IEEEauthorrefmark{1},
        Angelo Garofalo\IEEEauthorrefmark{1},\\
        Luca Benini\IEEEauthorrefmark{1}\IEEEauthorrefmark{2},
        Francesco Conti\IEEEauthorrefmark{1},
        Davide Rossi\IEEEauthorrefmark{1}
        }

\IEEEauthorblockA{\IEEEauthorrefmark{1}Department of Electrical, Electronic and Information Engineering (DEI), University of Bologna, Italy}
\IEEEauthorblockA{\IEEEauthorrefmark{2}IIS Integrated Systems Laboratory, ETH Zurich, Switzerland}}

\maketitle
\begin{abstract}
The emerging trend of deploying complex algorithms, such as Deep Neural networks (DNNs), increasingly poses strict memory and energy efficiency requirements on Internet-of-Things (IoT) end-nodes. Mixed-precision quantization has been proposed as a technique to minimize a DNN's memory footprint and maximize its execution efficiency, with negligible end-to-end precision degradation. In this work, we present a novel hardware and software stack for energy-efficient inference of mixed-precision Quantized Neural Networks (QNNs). We introduce Flex-V, a processor based on the RISC-V Instruction Set Architecture (ISA) that features fused \textit{Mac\&Load} mixed-precision dot product instructions; to avoid the exponential growth of the encoding space due to mixed-precision variants, we encode formats into the Control-Status Registers (CSRs). Flex-V core is integrated into a tightly-coupled cluster of eight processors; in addition, we provide a full framework for the end-to-end deployment of DNNs including a compiler, optimized libraries, and a memory-aware deployment flow. Our results show up to 91.5 MAC/cycle and 3.26 TOPS/W on the cluster, implemented in a commercial 22nm FDX technology, with up to 8.5$\times$ speed-up, and an area overhead of only 5.6\% with respect to the baseline. To demonstrate the capabilities of the architecture, we benchmark it with end-to-end real-life QNNs, improving performance by 2$\times$ - 2.5$\times$ with respect to existing solutions using fully flexible programmable processors.
\end{abstract}


\section{Introduction and Related Work} 
\label{intro}
Modern IoT applications require end-nodes to acquire raw data from sensors, extract ``distilled'' high-level features by applying near-sensors analytics including state-of-the-art ML and DL algorithms, and transmit this semantically dense information to higher-level nodes through wireless channels. 
However, running these models on embedded microcontroller systems poses severe challenges due to limited on-chip memory, power budget, and compute capabilities, requiring optimizations on both hardware and software.

A well-established solution to shrink the full-precision DL models to fit the limited storage available on microcontrollers is the adoption of low-bitwidth (8-bit or less) integer arithmetic to represent their parameters, after post-training quantization~\cite{qnn-mixed} or quantization-aware training~\cite{hubara}. These techniques have been demonstrated on state-of-the-art DNN topologies, adopting uniform or mixed-precision quantization schemes, reducing the model footprint by 47\% with a Top-1 accuracy drop in the range of 3.4\%, without significant impact upon the user experience of many IoT applications. Banner~et~al.~\cite{banner} propose a post-training 4-bit quantization method with an accuracy drop of a few percent, while authors of~\cite{qnn-mixed} presented further improvements reducing the memory footprint of DNNs up to 7$\times$ at the cost of an accuracy drop of only 4\%.

If well supported by the hardware processing systems, reduced precision integer arithmetic offers a significant efficiency boost with respect to floating-point operations. Low-bitwidth integer formats are widely adopted in custom digital and analog accelerators such as UNPU~\cite{unpu}, supporting fully-variable 1 to 16 bit weight bit-precision and delivering a peak energy efficiency of 50.6 TOPS/W at a throughput of 184 GOPS.  
Emerging Analog in-Memory Computing (AiMC) accelerators such as DIANA~\cite{diana} also implicitly exploit quantization, delivering peak energy efficiency in the range of 100-1000 TOPS/W. However, the high performance and efficiency of hardwired accelerators are counterbalanced by their poor flexibility, which makes it hard to deploy real-sized end-to-end DNNs on these systems and to achieve actual efficiencies similar to the theoretical peak. Limited flexibility and high area cost per device make them hard to adopt in IoT applications.

A compromise solution between dedicated accelerators and fully programmable devices is represented by FPGAs, where embedded general-purpose processors are coupled to the DSP-capable hardware to accelerate DNNs~\cite{qiu}. Several works explore reduced-precision arithmetic~\cite{finn}, but within a power envelope orders-of-magnitude larger than IoT nodes budget. Lattice proposed the SensAI stack~\cite{sensai}, which offers machine learning ultra-low-power (1 mW to 1 W) capabilities on FPGAs. However, these solutions have a limited number of LUTs and a non-negligible unit cost, not compatible with many IoT applications.
Hardware reconfigurability of these platforms offers higher flexibility than ASICs, but it is still far from the average IoT programmer demand. Additionally, their efficiency is much lower than that of ASICs.

The highest flexibility for QNN inference is offered by commercial general-purpose processors coupled with optimized software libraries such as CMSIS-NN libraries \cite{cmsis-nn} for ARM Cortex M4 and M7 processors.
A recent approach to enhance the computing capabilities of low-power MCU systems is through domain-specific Instruction Set Architecture (ISA) extensions. To address the DNN computing at the extreme edge, ARM presented the Cortex M-55 core based on the ARMv8-1M ISA, including an M-Profile Vector Extension (MVE) called Helium~\cite{ARMHELIUM} that also supports 8-bit MAC instructions. Unfortunately, microcontrollers implementing this ISA are not yet commercially available. Many solutions in the RISC-V ecosystem leverage this approach as well; for example, authors of~\cite{ri5cy} propose the XpulpV2 custom RISC-V ISA extensions for DSP applications, including support for 16-/8-bit SIMD operations. However, this ISA incurs performance degradation on sub-byte or mixed-precision linear kernels, since additional extra-instructions are required for data manipulation, introducing huge overhead~\cite{cmix-nn, pulp-nn-mixed}.

To boost sub-byte uniform linear kernels, XpulpNN~\cite{xpulpnn} extends XpulpV2 with 4- and 2-bit SIMD operations. Moreover, it introduces fused \textit{Mac\&Load} instructions that allow concurrent execution of SIMD dot-product operations with memory accesses, increasing the computation efficiency of the core up to 94\%. XpulpNN outperforms the performance of commercially available Cortex-M cores by up to 20$\times$ on quantized DNN layers. However, when operating on mixed-precision inputs, the efficiency boost of XpulpNN narrows significantly because of the massive software overhead necessary for packing and unpacking data. To eliminate the performance degradation compared to uniform precision kernels, authors of~\cite{mpic} propose direct hardware support for mixed-precision operations with dedicated RISC-V ISA extensions. To reduce the number of mixed-precision instructions to be encoded into the ISA, they exploit the \textit{dynamic bit-scalable execution mode}: the ISA instruction only encodes the type of the operation, while its format is specified by a Control Status Register (CSR) of the core.

In this work, we present a new hardware and software stack targeting energy-efficient inference of mixed-precision QNNs on a parallel cluster of RISC-V processors. The main contributions of this paper are the following:
\begin{itemize}
    \item We extend the RISC-V ISA with fused mixed-precision \textit{Mac\&Load} instructions. The proposed instruction set extension allows to achieve an ASIC-like utilization of MAC units in the cores (larger than 80\%), being able to operate on all the mixed-precision variants. Considering the mixed-precision capabilities and the preserved flexibility for general-purpose applications, we name our processor Flex-V.
    \item We integrate the extended processor in an eight-cores parallel ultra-low-power (PULP) cluster implemented in a commercial 22nm FDX technology to evaluate accurately the impact of such extensions on the operating frequency, area, and power.
    \item We integrate the proposed hardware extensions in a software framework for the end-to-end deployment of DNNs including a compiler, optimized libraries, and a memory-aware deployment flow, and we compare the proposed solution with the state-of-the-art end-to-end DNNs.
\end{itemize}

\begin{table}[t]
\centering
\scriptsize
\caption{Overview of QNN Embedded Computing Platforms and Main Metrics}
 \label{tab: related_work}
  \begin{tabular}{ccccc}
     %\toprule
     \hline
                                                    & Throughput    & Energy        & Power   & Flexibility \\
                                                    & [Gop/s]       & Efficiency    & Budget  &             \\
                                                    &               & [Gop/s/W]     & [mW]    &             \\
     \hline 
     \textbf{ASICs \cite{unpu}}                     & 1K - 50K      & 10K - 100K    & 1 - 1K  & Low         \\
     \hline
     \textbf{FPGAs \cite{sensai}}                   & 10 - 200      & 1 - 10        & 1 - 1K  & Medium      \\
     \hline
     \textbf{MCUs \cite{pulp-nn-mixed}}             & 0.1 - 2       & 1 - 50        & 1 - 1K  & High        \\
     \hline
     \textbf{This Work}                             & 25 - 85       & 610 - 3K      & 1 - 100 & High        \\
    \hline
\end{tabular}
\vspace{-0.5cm}
\end{table}

We compare the extended processor with state-of-the-art architectures by running both single convolutional layers and full end-to-end QNNs, we report a summary in Table \ref{tab: related_work}. Our results show a performance improvement, with respect to the execution with the extensions disabled, up to 8.5$\times$ on a single layer and up to 2.5$\times$ on the end-to-end network with negligible degradation of accuracy and a peak energy efficiency of 3.26 TOPS/W, approaching that of accelerators, at low area cost 5.6\% with respect to the baseline processor cluster and without compromising flexibility. The hardware and software described in this work are open-source, to support and boost an innovation ecosystem focusing on ultra-low-power computing for the IoT landscape.

\section{Background} \label{background}
% PULP cluster
\subsection{PULP cluster}
%\noindent
Parallel Ultra-Low Power (PULP) is an open-source computing platform exploiting near-threshold computing to reach high energy efficiency, leveraging parallelism to enhance the performance degradation at low voltage \cite{vega}.
% PULP system
\begin{figure}%[h]
    \centering
    \includegraphics[width=\linewidth]{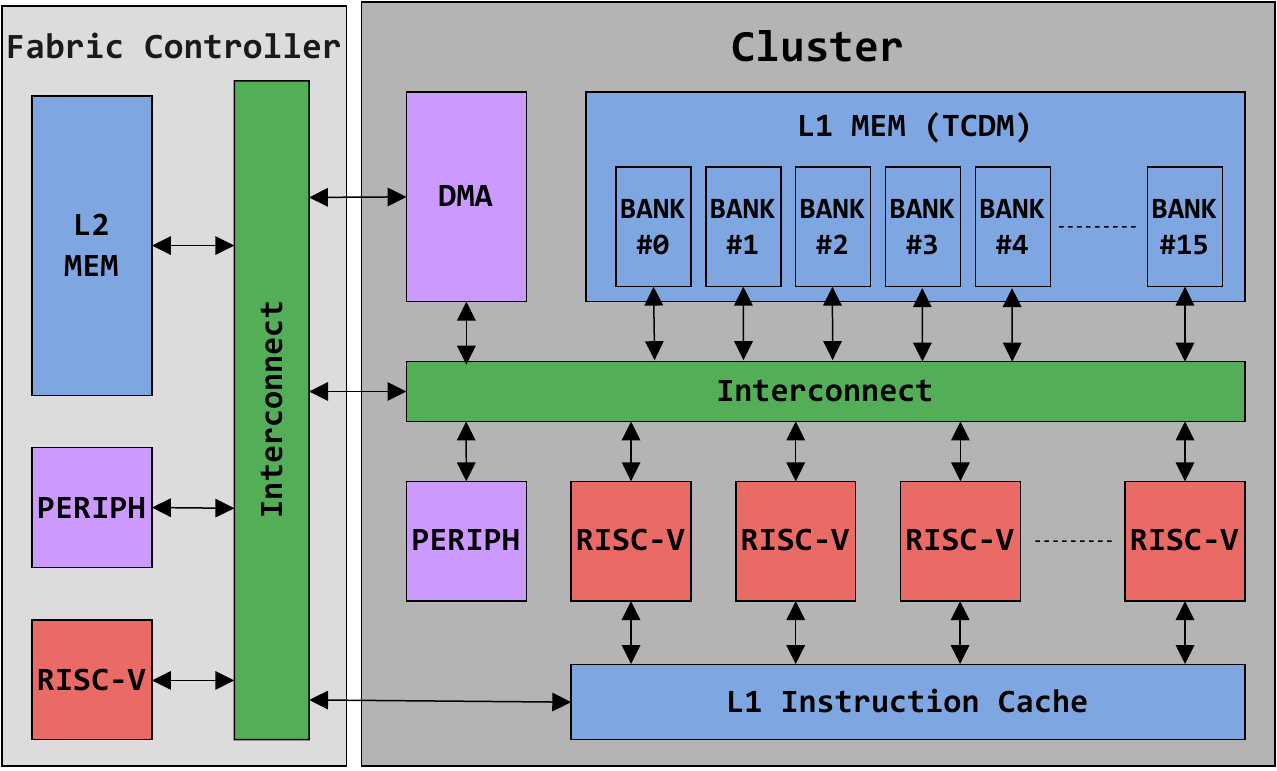}
    \caption{Parallel Ultra-Low Power (PULP) system, consisting of a Fabric Controller (FC) accelerated by a parallel cluster of 8 RISC-V based processors.}
    \label{fig: pulp}
    \vspace{-0.5cm}
\end{figure}
The PULP cluster adopted as a reference, shown in Fig.~\ref{fig: pulp}, is composed of eight RI5CY cores~\cite{ri5cy}, each of them characterized by a 4-stage in-order single-issue pipeline and the RV32IMCXpulpV2 Instruction Set Architecture (ISA). XpulpV2 is a specialized extension to the RISC-V ISA~\cite{ri5cy} designed for efficient digital signal processing (DSP) computation. It features hardware loops, post-modified access load and store instructions, along with SIMD operations on 16-bit and 8-bit integer vector operands.

The cores in the cluster share data on a Tightly Coupled Data memory (TCDM) of 128 kB, divided into 16 banks. The memory is accessed through a one-cycle latency logarithmic interconnect. The PULP cluster accelerator and its host, i.e. the Fabric Controller (FC), communicate through an AXI interface. Data transfers between the TCDM and the second level of memory, hosted by the Fabric Controller, are managed by a dedicated DMA.

\begin{figure*}
    \centering
    \includegraphics[width=\linewidth]{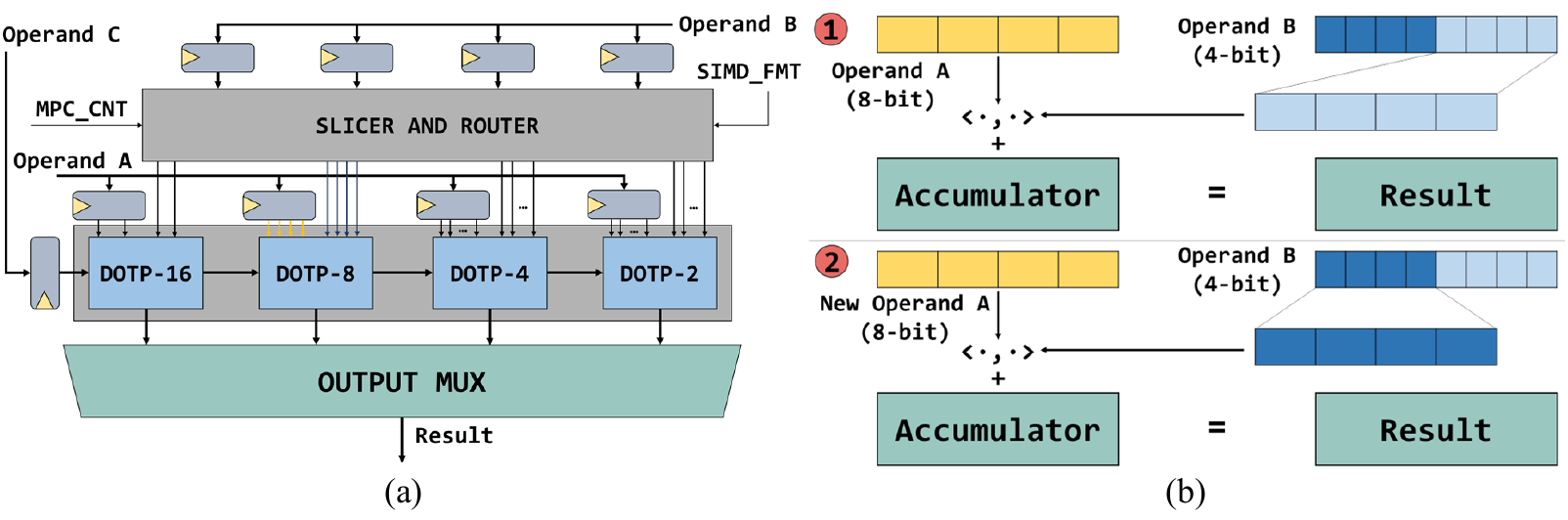}
    \caption{\textit{(a)} Dotp Unit and \textit{(b)} execution flow of a mixed-precision \textit{sumdotp} instruction between 8-bit operand A and 4-bit operand B.}
    \label{fig: dotp}
    \vspace{-0.5cm}
\end{figure*}

The cluster processors fetch the instructions from a two-levels (the first private to each core, the second shared) hierarchical instruction cache to enhance the hit rate. The cluster is also provided with the Hardware Synchronization Unit which manages fine-grained parallel thread dispatching and clock-gating of idle cores waiting for synchronization, enabling low-overhead and fine-grained parallelism, thus high energy efficiency.

\subsection{QNN execution model} \label{matmul worload}

The software stack we propose in this work extends the PULP-NN software library presented in~\cite{pulp-nn-mixed}. It relies on the Height-Width-Channel (HWC) data layout and on an execution model optimized for resource-constrained microcontrollers. Convolution layers are implemented by combining three distinct phases:

\textit{\textbf{im2col:}} for a given output pixel position, the 3D input activations (in HWC format) of the current convolution are re-arranged into a 1D vector along the filter and input channel dimensions. PULP-NN performs this operation simultaneously for 2 output pixels, producing two separate \textit{im2col buffers}.

\textit{\textbf{Matrix Multiplication:}} this step consists of a \textit{sum-of-dot-product} operation between the current \textit{im2col} buffer and the sets of filters to produce the intermediate outputs at the higher 32-bit precision. 
The kernel leverages the XpulpV2 ISA and exploits data locality within the Register File (RF) of RI5CY to maximize the computation throughput. As a result of design exploration in the space of registers resources available in the RI5CY RF, it is possible to implement a MatMul with a “$4\times2$" unrolling factor, fetching from memory the weights from two consecutive filters and the input activations from two different \textit{im2col} buffers to produce two activation outputs related to four consecutive channels, in the same inner loop of the MatMul. 

%\noindent
\textit{\textbf{Quantization:}} each intermediate 32-bit accumulator from the previous stage is represented back in low-bitwidth form by applying normalization and quantization functions, composed of one MAC, one shift, and one clip operation.

\section{Flex-V core Architecture} \label{architecture}
We introduce Flex-V, our RISC-V ISA-extended processor that flexibly supports sub-byte and mixed-precision preserving the fully-programmable capabilities of a general-purpose processor. The three key concepts that guided the development of the core's micro-architecture are \textit{dynamic bit-scalable} execution, \textit{fused Mac\&Load}, and \textit{fully-flexible mixed-precision}. In \textit{dynamic bit-scalable} execution, a particular op-code defines a whole family of \textit{virtual instructions}; the choice of a particular one to execute depends on contextual bits stored in a status register of the core. 

\begin{figure}[tb]
    \centering
    \includegraphics[width=0.9\linewidth]{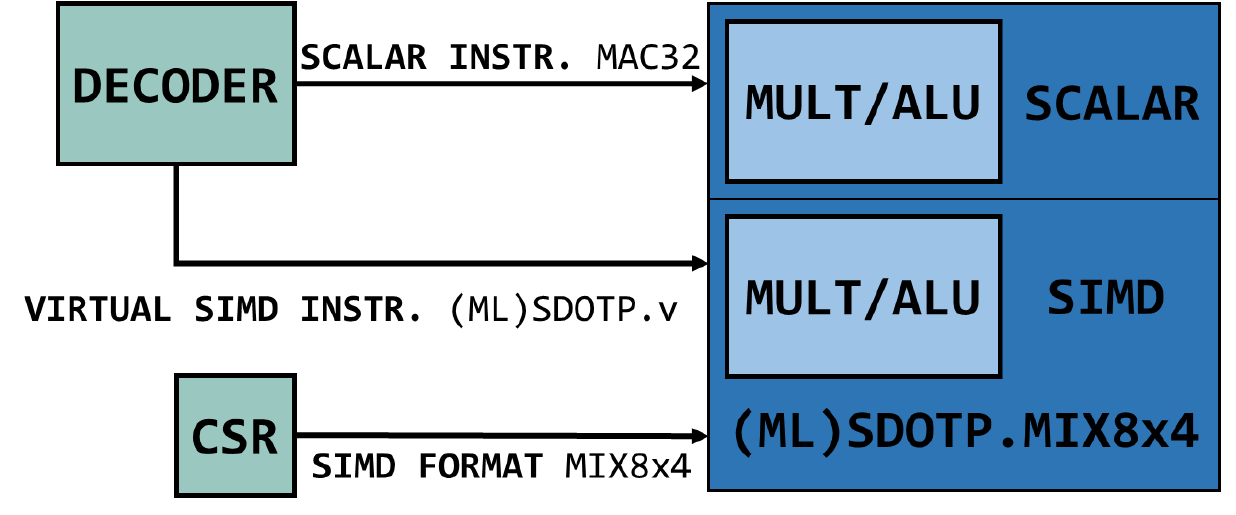}
    \caption{Instruction Decoding during status-based execution.}
    \label{fig: status-based}
    \vspace{-0.5cm}
\end{figure}

Fig. \ref{fig: status-based} shows the decoding process when the Flex-V core is running in \textit{dynamic bit-scalable} execution mode: in case of a \textit{Scalar} instruction, the decoder extracts all necessary information from the encoding of the instruction and communicates it to the EX stage. Contrarily, if the received op-code corresponds to a \textit{Virtual SIMD} instruction, e.g. a \textit{(ml)sdotp}, the decoder enables the proper functional unit within the EX stage, but the precision of the operation's operands depends also on status bits stored in the CSRs, such as the SIMD format, and on signals coming from dedicated controllers.

% VIRTUAL INSTRUCTION

The mixed-precision Dot Product (Dotp) Unit shown in Fig. \ref{fig: dotp}a, exploiting the operand-precision information stored in the CSRs, implements the sub-byte and mixed-precision support. It integrates dedicated units for 4- and 2-bit operands together with a \textit{Slicer\&Router} responsible for their extraction from a 32-bit input word. Considering the \textit{sum-of-dot-product (sdotp)} operation between an 8-bit operand A and a 4-bit operand B, only four elements within the 32-bit input word for B can be consumed by a single instruction: as shown in Fig.~\ref{fig: dotp}b, the Slicer selects either the first or the last four elements depending on the value of \textit{MPC\_CNT} signal, then the Router directs the selected elements to the Dotp sub-unit specified by the \textit{SIMD\_FMT} signal coming from the CSR, i.e. the DOTP-8 for this example. The overall process is governed by the Mixed-Precision Controller (MPC).

\begin{figure}
    \centering
    \includegraphics[width=\linewidth]{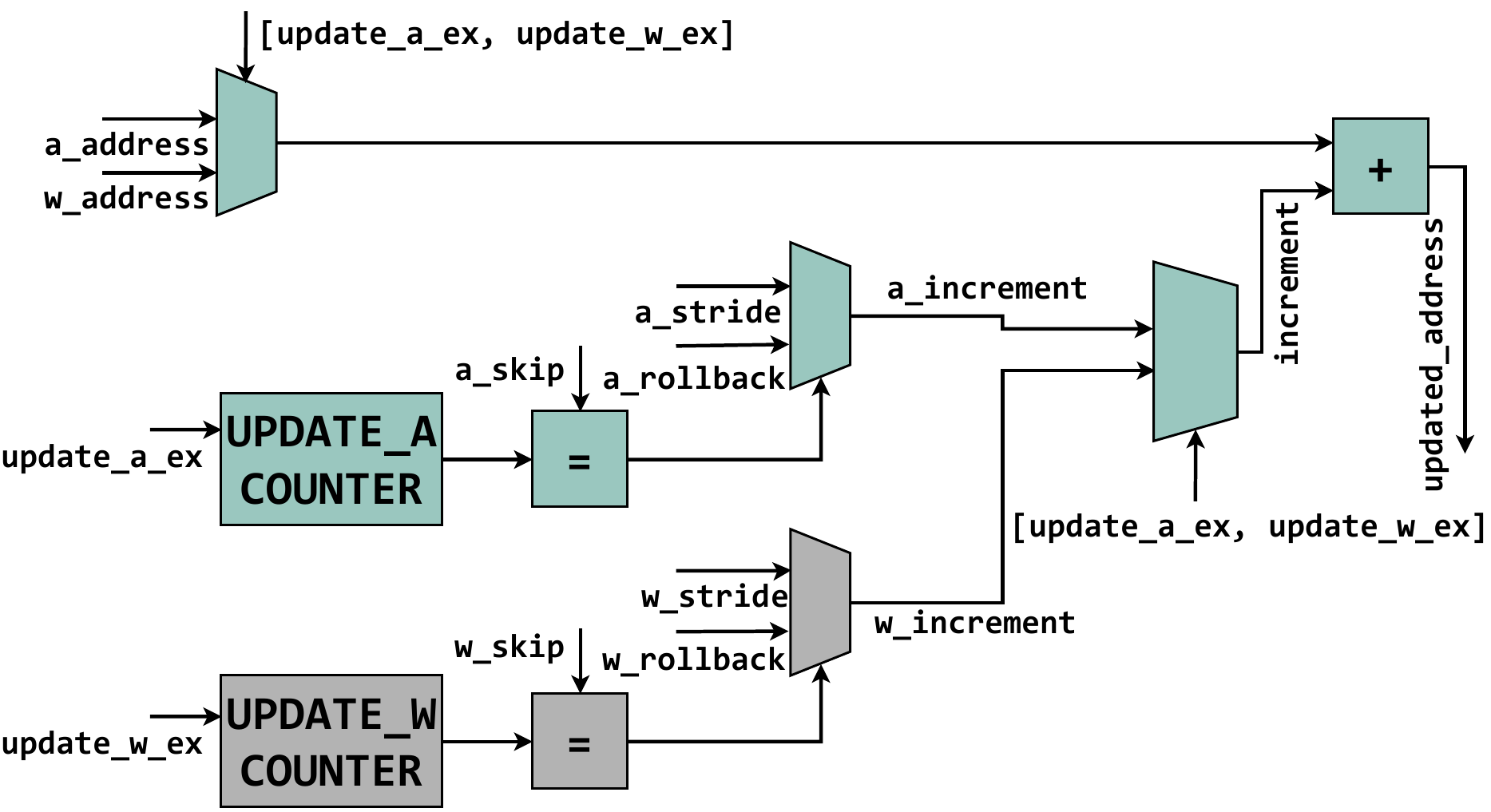}
    \caption{Functional schematic of the \textit{Mac\&Load} Controller. Active blocks during an activation update are highlighted in green.}
    \label{fig: macload controller}
    %\vspace{-0.5cm}
\end{figure}

%  MAC & LOAD: NN-RF and MLC
The \textit{fused Mac\&Load (mlsdotp)} instruction overlaps a SIMD \textit{dot-product}-like operation with a Load performed during the writeback stage, typically to replace non-stationary data in a register feeding the following \textit{Mac\&Load} instruction. Finally, we define \textit{fully flexible mixed-precision} as the hardware support for automatic management of instructions involving operators with different bit precision. An additional Neural Network Register File (NN-RF), with six 32-bit registers dedicated to values of activations and weights, has been added to enable Load operations during the \textit{Mac\&Load} write-back stage, which cannot be done in the general purpose register file (GP-RF). Finally, the core includes a \textit{Mac\&Load} Controller (MLC) that is in charge of the automatic address generation, described in Fig. \ref{fig: macload controller}.

% MLC

\begin{figure}
    \centering
    \includegraphics[width=0.9\linewidth]{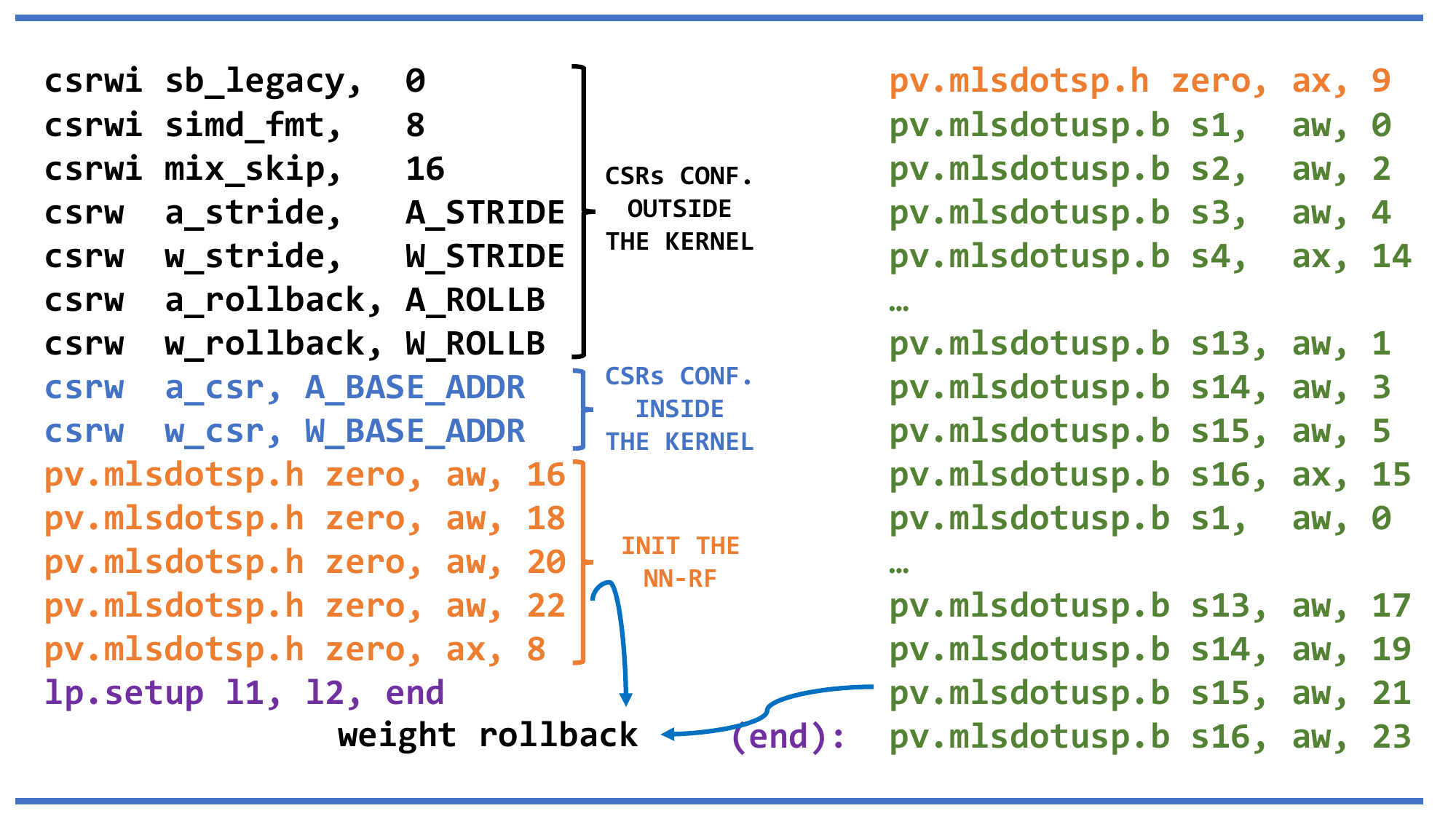}
    \caption{Pseudo-assembly code of a “4$\times$4" \textit{MatMul} between 8-bit activations and 4-bit weights.}
    \label{fig: mm_assembly}
    \vspace{-0.5cm}
\end{figure}

Fig. \ref{fig: mm_assembly}, shows an assembly snippet of a Matrix Multiplication kernel between 8-bit activations and 4-bit weights. The kernel starts initializing the CSRs driving the inputs of the MLC (i.e. 
\textit{\{w,a\}\_skip},
\textit{\{w,a\}\_stride}, \textit{\{w,a\}\_rollback}), and the MPC parameters defining encoded activations and weight precision (\textit{simd\_format}) and the weight reuse parameter (\textit{mix\_skip}). Once the CSRs needed to configure MLC and MPC are set, the inner loop of the kernel starts the execution. After the initialization of the base memory addresses, four weights and one activation are loaded explicitly to fill the NN-RF. The innermost loop executes only one explicit load per iteration, then all other updates of the NN-RF are performed in the writeback phase of the \textit{Mac\&Load} instruction. Strides, rollbacks, and thresholds are all stored in CSRs and they depend only on static features of the MatMul, such as the number of input channels, the dimensions of the filter kernel, and the precision of the operands.

\begin{figure}%[bt]
    \centering
    \includegraphics[width=0.98\linewidth]{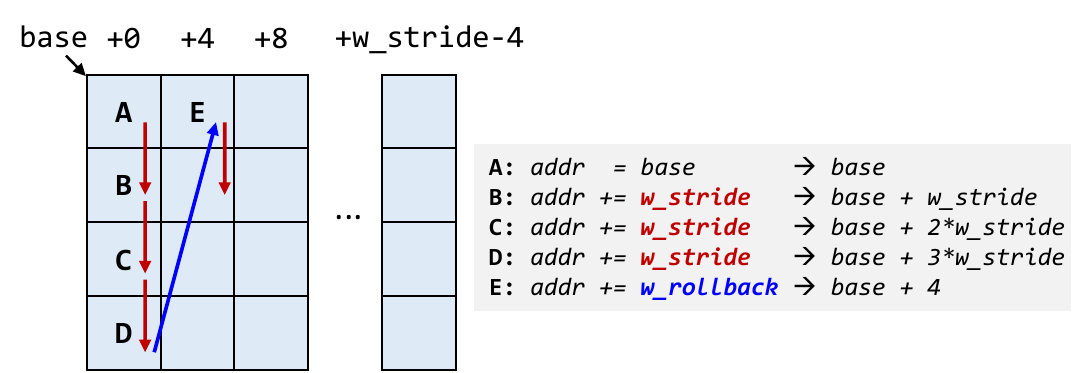}
    \caption{Example of regular addressing pattern for \textit{w} in  a  MatMul with unrolling factor “4$\times$2".}
    \label{fig:pointers}
    \vspace{-0.5cm}
\end{figure}

% 
% MLC DETAILS
During the execution of the inner loop the MLC automatically generates the memory address for both operands: it navigates a two-dimensional strided pattern by updating a register-stored pointer \textit{\{w,a\}\_addr} with three static parameters, \textit{\{w,a\}\_stride}, \textit{\{w,a\}\_rollback}, and \textit{\{w,a\}\_skip}. The \textit{\{w,a\}\_stride} parameter corresponds to the stride in the direction of the innermost loop of the pattern, while \textit{\{w,a\}\_rollback} rolls back the pointer of all innermost loop iterations and adds the stride of a single outermost loop iteration.
\textit{\{w,a\}\_skip} is the number of innermost loop iterations. Fig.~\ref{fig:pointers} shows the pattern in the example of a MatMul with unrolling factor “4$\times$2" in PULP-NN~\cite{pulp-nn-mixed}.  This kind of pattern would require substantial instruction overhead ($\sim$30\%) for pointer management; the MLC deals with this entirely in hardware. 

We can also note that, in case of mixed-precision inputs, there's an additional degree of unrolling with respect to uniform precision execution: thanks to the hardware support for mixed-precision, each 32-bit register dedicated to weights can be exploited from 2 to 4 times to process different activations, then they're updated at the end of each innermost loop iteration. These features, together with the automatic update of the activations and weights pointers enabled by the MLC, increase the utilization of MAC units reducing the overall number of loads from memory. Furthermore, it can be noted that, while the baseline core (i.e. RI5CY) is limited to a maximum unrolling factor of “4$\times$2" that saturates the registers within the GP-RF, the introduction of the dedicated NN-RF in Flex-V extends it to “4$\times$4" further improving data reuse, hence performance.

\section{Deployment Flow}\label{deployment}
We develop an optimized software library to take advantage of the proposed ISA extensions, replacing software-based low-precision data unpacking with the hardware support for sub-byte and mixed-precision operands, and introducing the new unrolling degree for matrix multiplications between mixed-precision operands.

To deploy end-to-end real-sized QNN benchmarks, we extend the open-source DORY tool~\cite{dory}\footnote{\texttt{https://github.com/pulp-platform/dory}} to support low-precision data formats ($<$ 8-bit). 
The tool automatically produces template-based C code that wraps a target backend, managing different levels of memories (i.e., L1, L2, and the external RAM) and orchestrating the tensor movements.
In particular, DORY exploits a tiling approach to separate layers into small nodes whose tensors can fit the L1 memory of the system. Then, it produces C routines which \textit{i)} execute these smaller nodes in L1, and \textit{ii)} double-buffer the movements of tensors from L2 to L1. 
Notice that since the DMA is not blocking, the calls to the kernels are always overlapped with the asynchronous DMA calls.

The existing tool only supported 8-bit integer tensors. 
To plug in our new library, we modify the key elements of DORY to support 2-bit and 4-bit data formats.
First, we extend the tiling solver based on Constraint Programming to support different data formats: now it considers the new constraints associated with sub-byte data formats, i.e., that the convolutional loop's innermost dimensions should always be byte-aligned.
Then, we create a new set of templates to support the new ISA extensions. In the new templates, before the tiling loops, we insert the CSRs setup that is common to every sub-nodes executed. Inside the tiling loops, we call the functions implementing the key kernels exploiting the new proposed ISA extensions.
Finally, we adjust the DORY mapping tool to consider that layers' tensors can have different data formats, correctly sizing the data transfers between L3, L2, and L1.

\section{Results and Discussion}
To evaluate the Flex-V core in terms of timing, power, and area overhead compared to other cores based on the RI5CY architecture, we integrate RI5CY, MPIC, XpulpNN and Flex-V cores into the PULP cluster and perform separate full implementations with the Global Foundries 22nm FDX technology node. To evaluate the proposed hardware-software stack, we benchmark the PULP cluster with the Flex-V cores on synthetic convolutional layers and on the full deployment of real-world end-to-end QNNs.

\subsection{Physical Implementation}
We synthesize the PULP clusters with Synopsys Design Compiler-2019.12 and perform full place\&route flow with Cadence Innovus-17.11.000 using the worst-case corner (SSG 0.59V, -40$\degree$C/125$\degree$C). To perform power overhead evaluations between RI5CY and Flex-V with disabled extensions, we run timing-annotated post-layout simulations of 8-bit MatMuls in typical corners at 250 MHz.

The total area of the Flex-V core is 0.018 mm$^2$, with an overhead of 30\% compared to RI5CY due to the additional logic to extend the Dotp Unit and implement the MLC and the MPC. We note that the impact is only 6\% when we compare the area at the cluster level. The additional logic of Flex-V compared to RI5CY does not significantly impact the maximum operating frequency of the cluster~(-2\%), which peaks up to 463 MHz.

Note that, despite the additional logic introduced to implement the new ISA extensions, the power consumption overhead with respect to RI5CY related to the execution of an 8-bit MatMul with only XpulpV2 extensions is limited to 2.47\% for the single processor and 2.04\% for what concerns the whole cluster thanks to clock-gated CSRs. Complete area and power results are reported in Table~\ref{tab: area-power}.

\begin{figure*}[tb]
    \centering
    \includegraphics[width=\linewidth]{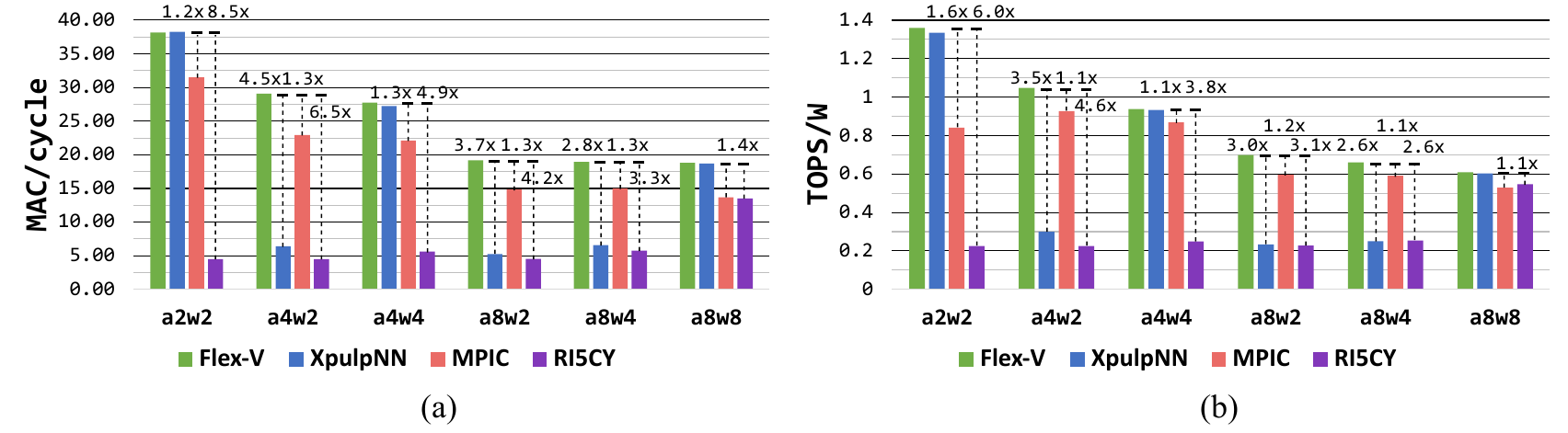}
    \caption{Performance \textit{(a)} and energy efficiency \textit{(b)} of single convolutional kernels executed on the PULP cluster.}
    \label{fig: perf-eff}
    \vspace{-0.5cm}
\end{figure*}

\begin{table}[tb]
    \centering
    \scriptsize
    \caption{Area and Power Consumption Results}
    \begin{tabular}{c|c|c}
         \hline
         %\hline
                                    & \textbf{RI5CY}   & \textbf{Flex-V} \\
         \hline
         %\hline
         \textbf{$f_{max}$ [MHz]}   & 472              & 463 \\
         \hline
         %\hline
         \multicolumn{3}{c}{\textbf{Area [$\mu\textrm{m}^2$] (Overhead w.r.t. baseline [\%])}} \\
         \hline
         \textbf{Cluster}           & 518227           & 547211 (5.59 \%) \\
         \hline
         \textbf{Core}              & 13721            & 17816  (29.8 \%)\\
         \hline
         %\hline
         \multicolumn{3}{c}{\textbf{Core Power [mW] 8b MatMul (Overhead w.r.t. baseline [\%])}} \\
         \hline
         \textbf{Dynamic}           & 0.160            & 0.162 (0.87 \%)\\
         \hline
         \textbf{Leakage}           & 0.024            & 0.037 (56.8 \%)\\
         \hline
         \textbf{Total}             & 0.825            & 0.846 (2.47 \%) \\
         \hline
         %\hline
         \multicolumn{3}{c}{\textbf{Cluster Power [mW] 8b MatMul (Overhead w.r.t. baseline [\%])}} \\
         \hline
         \textbf{Dynamic}           & 2.34             & 2.37 (1.32 \%)\\
         \hline
         \textbf{Leakage}           & 0.613            & 0.71 (15.7 \%)\\
         \hline
         \textbf{Total}             & 12.3             & 12.6 (2.04 \%) \\
         \hline
         %\hline
    \end{tabular}
    \label{tab: area-power}
    \vspace{-0.2cm}
\end{table}

\subsection{DNN Layers Benchmarking}
\label{sec:benchmarking}
To demonstrate the benefits of the proposed core, we benchmark the PULP cluster with Flex-V cores on a set of synthetic convolution kernels, in terms of performance and energy efficiency, and we compare it with RI5CY\cite{ri5cy}, MPIC\cite{mpic} and XpulpNN\cite{xpulpnn}. The layers operate on representative tiles used in such types of devices to deploy QNN inference, applying 64$\times$3$\times$3$\times$32 filters on a 16$\times$16$\times$32 input tensor and featuring different bit-precision for activations and weights (including mixed-precision). The results are then compared against the cluster execution of the same kernels on similar RISC-V cores and reported in Fig.~\ref{fig: perf-eff}.  
Although MPIC~\cite{mpic} supports mixed-precision operations in its ISA, our solution speeds-up convolution kernels by 1.4$\times$ thanks to the Mac-Load mechanism available in the Flex-V core and the supported “$4\times4$" MatMul format. Moreover, the performance boost of Flex-V grows up to 4.5$\times$ and 8.5$\times$ with respect to XpulpNN and XpulpV2, respectively, which show heavy performance degradation on mixed-precision and sub-byte QNN kernels due to lack of support in the ISA for these operations that require adding extra-instructions in the assembly for data manipulation.

Table \ref{tab: matmul} shows that the proposed architecture reaches a peak of energy efficiency of 3.26 TOPS/W on the uniform 2-bit MatMul kernel and 870 GOPS/W on the 8-bit configuration, which is comparable to dedicated hardware acceleration units without giving away software flexibility. Flex-V outperforms all the other solutions for all the configurations.

\begin{table}[tb]
    \centering
    \scriptsize
    \caption{Performance [MAC/cycle] / Energy Efficiency [TOPS/W] of MatMul kernels}
    \begin{tabular}{c|c|c|c|c}
        \hline
        \textbf{Inputs} & \textbf{RI5CY}    & \textbf{MPIC} & \textbf{XpulpNN}  & \textbf{Flex-V} \\
        \hline
        \textbf{a2w2}   &  -                & 57.44 / 0.84  & 90.8 / 2.99      & 91.5 / 3.26  \\
        \hline
        \textbf{a4w2}   &  -                & 35.91 / 0.93  & 7.62  / 0.23      & 51.9 / 1.87  \\
        \hline
        \textbf{a4w4}   &  -                & 32.08 / 0.87  & 49.5 / 1.60      & 50.6 / 1.71  \\
        \hline
        \textbf{a8w2}   & 4.91 / 0.25       & 19.55 / 0.60  & 6.07  / 0.20      & 27.8 / 1.01  \\
        \hline
        \textbf{a8w4}   & 6.38 / 0.28       & 19.19 / 0.59  & 7.63  / 0.20      & 27.6 / 0.96  \\
        \hline
        \textbf{a8w8}   & 16.6 / 0.67       & 16.45 / 0.53  & 26.1 / 0.79      & 26.9 / 0.87  \\
        \hline
    \end{tabular}
    \label{tab: matmul}
    \vspace{-0.5cm}
\end{table}

\subsection{End-to-end Networks}

To further demonstrate the capabilities of the proposed architecture, we benchmarked it with end-to-end real-life QNNs exploiting the deployment flow described in Section~\ref{deployment}. We considered three use cases: an 8-bit MobileNetV1, a fully mixed-precision 8b4b MobileNetV1 and an aggressively quantized 4b2b ResNet-20. The two MobileNetV1 networks have been trained on ImageNet while the 4b2b ResNet-20 targets CIFAR10. It is worth noticing that the memory footprint of the 8b4b MobileNetV1 is reduced by 47\% with respect to the 8-bit quantized model while its accuracy reaches 66.0\%, with a degradation of only 3.3\%. Performance and accuracy of all the tested networks are reported in Tab.~\ref{tab: benchmark nets}: the experiments performed on the ResNet-20 featuring 4-bit activations and 2-bit weights show that the proposed architecture achieves 2.3$\times$ and 2.5$\times$ of speedup with respect to XpulpV2 and XpulpNN. We also report the results of the execution of an end-to-end network on the STM32H7 presented by Capotondi et al.~\cite{cmix-nn}: the speedup of the proposed architecture with respect to this commercial product reaches 19$\times$ thanks to the combination of the extended ISA and the optimized software executed on the eight-core cluster.

\begin{table}[tb]
    \centering
    \caption{Accuracy, Memory Footprint and Perf. of end-to-end networks}
    \scriptsize
    \begin{tabular}{c|c|c|c}
        \hline
        \textbf{Network}    & \textbf{MNV1 (8b)} & \textbf{MNV1 (8b4b)}    & \textbf{ResNet20 (4b2b)} \\
        \hline
        \textbf{Top-1 Acc.}       & 69.3\%             & 66.0\% & 90.2\% ~\cite{hawq}\\
        \hline
        \textbf{Deg. w.r.t. 8b}    & -           & 3.3\%    & 0.15\%\\
        \hline
        \textbf{Model size}       & 1.9 MB             & 997 kB            & 142 kB \\
        \hline
        \textbf{Mem. saved}       &                    & 47\%              & 63\%\\
        \hline
        \multicolumn{4}{c}{\textbf{Performance [MAC/cycle]}} \\
        \hline
        \textbf{STM32H7}    & 0.33     & 0.30       & -\\
        \hline
        \textbf{XpulpV2}    & 5.6      & 3.2             & 4.8 \\
        \hline
        \textbf{XpulpNN}    & 6.0       & 2.7             & 4.4  \\
        \hline
        \textbf{Flex-V}  & 6.0      & 5.8             & 11.2 \\
        \hline
    \end{tabular}
    \label{tab: benchmark nets}
    \vspace{-0.5cm}
\end{table}

\section{Conclusion}
\noindent
We presented a novel hardware and software stack that meets the challenge of energy-efficient mixed-precision QNN inference on MCU processors core. We extended the RISC-V ISA with sub-byte and mixed-precision fused \textit{Mac\&Load} instructions aiming to remove the overhead caused by loading and unpacking data before actual computation. We integrated the Flex-V core, which implements the extended ISA, into a tightly-coupled PULP cluster of eight cores. Its implementation with GF22FDX technology shows an area overhead of only 5.6\% with respect to the baseline cluster with RI5CY cores. Furthermore, we developed a software library leveraging the new ISA extensions to improve the performance of convolutional kernels, key kernels to boost the execution of end-to-end QNNs. The results on single convolutional layers show up to 38.2 MAC/cycle boosting by 8.5$\times$ and 4.5$\times$ the execution on RI5CY and XpulpNN cores, respectively. We also benchmarked the proposed architecture with three end-to-end real-life QNNs, obtaining a performance gain of 2$\times$ - 2.5$\times$ with respect to state-of-the-art solutions. From the physical implementation of the cluster in 22nm FDX technology, we observed a peak energy efficiency of 3.26 TOPS/W.

\section*{Acknowledgement}
This work was supported in part by NeuroSoC HORIZON EU Project (g.a. 101070634) and in part by TRISTAN HORIZON EU Project (g.a. 101095947).

\bibliographystyle{IEEEtran}
\bibliography{bibliography.bib}

\end{document}